\documentstyle[twocolumn,epsf]{jpsj}

\title{
$d$- and $p$-wave superconductivity mediated by 
spin fluctuations 
in two- and three-dimensional single-band repulsive Hubbard model
}

\author{
Ryotaro Arita, Kazuhiko Kuroki and Hideo Aoki
}
\inst{Department of Physics, University of Tokyo, Hongo, Tokyo
113-0033, Japan}

\recdate{\today}

\abst{
We have systematically studied superconducting instabilities in the 
repulsive Hubbard model for $d$-wave and $p$-wave pairing 
in various 2D and 3D lattices.
Using fluctuation exchange approximation, 
we consider 3D face centered cubic lattice, 3D
body centered cubic lattice, 3D simple cubic lattice,
2D square lattice and 2D triangular lattice, 
where either strong ferro- or antiferromagnetic spin fluctuation
is present.
We show that (i) $d$-wave instability mediated by antiferromagnetic
spin fluctuations is stronger than $p$-wave instability
mediated by ferromagnetic spin fluctuations both in 2D and 3D, and
(ii)$d$-wave instability in 2D is much stronger than that in 3D.
These amount that the ``best" situation is the 
antiferromagnetic-fluctuation mediated in 2D 
as far as the single-band Hubbard model on ordinary lattices are concerned.
}

\kword{Hubbard model, anisotropic superconductivity,
fluctuation exchange approximation
}

\begin{document}
\sloppy
\maketitle

\def\runtitle{$d$- and $p$-wave superconductivity mediated by 
spin fluctuations}
\def\runauthor{Ryotaro Arita, Kazuhiko Kuroki and
Hideo Aoki}

\section{Introduction}
There has been an increasing fascination with 
electronic mechanisms for pairing 
since the high-$T_{\rm c}$ superconductivity was discovered in
the cuprates\cite{Bednorz}.
Among various mechanisms,
the pairing mediated by spin fluctuations has been 
proposed\cite{Moriya1,Moriya2,Moriya3,Pines1,Pines2,Pines3} 
and have been examined intensively.

The pairing interaction, denoted by $V^{(2)}$, 
resulting from exchanging spin fluctuations is 
dominated by the spin susceptibility which 
has a strong wavenumber dependence, so that 
$V^{(2)}$ in turn has a strong wavenumber dependence.  
This determines the symmetry of the pairing, since this dictates 
the dominant pair-hopping processes on the Fermi surface.
There, the dominant mode of spin fluctuations, 
hence the symmetry of the pairing, must be sensitively 
correlated with the lattice structure and filling in general.
For example, the respulsive electron correlation causes 
the face centered cubic (FCC) lattice with low density
of electrons to have strong ferromagnetic fluctuations, 
or the half-filled body centerd cubic (BCC) lattice to have
strong antiferromagnetic fluctuations as 
recently confirmed\cite{AriOnoKuAo}.
Thus, it is an interesting question to ask how the 
superconductivity would be correlated with the lattice 
structures.

As for the singlet, anisotropic ({\it e.g.}, d-wave) pairing 
in 2D systems, calculations on a microscopic level 
have been performed: For the repulsive Hubbard model, 
which is a simplest possible model of the electron correlation, 
the fluctuation exchange approximation (FLEX)
developed by Bickers {\it et al.},\cite{FLEX1,FLEX2} has been employed 
to show the occurrence of the $d$-wave superconductivity in 
a square lattice.\cite{Dahm,Deisz} 
As for other 2D lattices, several authors have suggested 
the occurrence of $d$-wave superconductivity in the 
Hubbard model on 2D anisotropic triangular lattice, which 
represents an organic superconductor (BEDT-TTF)$_2$X
\cite{KinoKontani,KonMori,Schmalian,VojtaDaggoto,Jujo1,Jujo2}.
These results indicate that 
$T_{\rm c}$ for the superconductivity near the antiferromagnetic 
instability in 2D is $\sim O(0.01t)$ ($t$: transfer integral), 
i.e., two orders of magnitude smaller 
than the original electronic energy, but still 
`high $T_{\rm c}$' $\sim O(100$ K) if we take $t\sim O(1$ eV) for the cuprates 
(while $T_{\rm c} \sim O(10$ K) if we take $t\sim O(0.1$ eV) for the organic 
conductors).  


As for 3D systems, Scalapino {\it et al}\cite{Scalapino}
showed that paramagnon
exchange near a spin-density wave instability gives rise to
a strong $d$-wave pairing interaction for the 3D Hubbard model
on the simple cubic lattice, 
but $T_{\rm c}$ was not discussed there.
Nakamura {\it et al}\cite{Nakamura} extended Moriya's spin fluctuation theory 
of superconductivity\cite{Moriya2} to 3D systems, 
and concluded that $T_{\rm c}$ is similar between the 2D and 3D cases 
provided that common parameter values 
(scaled by the band width) are taken. 
However, the parameters there are phenomelogical ones, 
so it is not clear whether the result remains valid 
for microscopic models.

The possibility of the triplet pairing, on the other hand,
has been investigated since the 1960's 
for superfluid $^3{\rm He}$\cite{Leggett}, a heavy
fermion system ${\rm UPt_3}$\cite{heavyFermion}, or most recently, an
oxide ${\rm Sr_2RuO_4}$\cite{Sr1,Sr2}.  The experimental results 
suggesting $p$-wave pairing in these materials have 
stimulated theoretical studies for electron-repulsion originated
superconductivity. 
For the electron gas model, Chubukov extended 
the Kohn-Luttinger theorem\cite{KohnLutt}, 
which asserts that the repulsively interacting electron gas should be 
instable against pairing formations at low enough temperatures, 
to $p$-wave pairing for 2D and 3D electron gas
in the dilute limit by analyzing the singularity of the scattering 
amplitude\cite{Chubukov1,Chubukov2}.  
Takada\cite{Takada} discussed the possibility of $p$-wave superconductivity
in the dilute electron gas with the Kukkonen-Overhauser model\cite{KO}.
This model considers the effective electron-electron interaction 
composed of the bare interaction and the interactions
mediated by charge/spin fluctuations that contains 
the so-called local-field correction. 

As for lattice systems, which is the subject of the present paper,
2D Hubbard model with large enough 
next-nearest-neighbor hopping $(t')$ has been shown to exhibit 
$p$-pairing for dilute band fillings.\cite{ChubukovLu} 
Hlubina\cite{Hlubina99} reached a similar conclusion by evaluating 
the superconducting vertex in a perturbative way.\cite{Takahashi}
However, the energy scale of the $p$-pairing in the Hubbard model, 
i.e., $T_{\rm c}$, has not been questioned so far.  
Using a phenomenological approach, 
Monthoux and Lonzarich\cite{MonLon} have recently concluded 
for 2D systems that the $d$-wave pairing is much stronger 
than $p$-wave pairing.  

Given these backgrounds, from microscopic view of point,
we investigate in this paper 
various 2D and 3D lattice structures 
(square, triangular, simple cubic, BCC, FCC)
with systematically varied next-nearest neighbor hopping 
to tune the dispersion and varied band filling.  
Specifically, we address the following fundamental questions:

\noindent (i) Can the pairing instability in 3D be 
stronger than that in 2D?

\noindent (ii) Can the pairing instability with other 
symmetry, i.e., spin-triplet $p$-pairing in the presence of 
{\it ferromagnetic} spin fluctuations, become strong?

A part of the present study has been briefly reported\cite{AKA-PRB}, 
while here we extend the calculation to various cases
in order to extensively confirm our previous conclusions.
Namely, we study possibility of $d$-wave superconductivity
for nearly half-filled square, simple cubic (SC), and BCC lattices 
where strong antiferromagnetic fluctuations are present, 
along with possibility of $p$-wave superconductivity
for low density
square lattice with significant next nearest-neighbor hopping, 
quarter-filled triangular lattice,
and low-density FCC lattice 
where ferromagnetic spin fluctuations should be dominant.

We employ the FLEX approximation, which enables us 
to handle strong spin fluctuations.
As seen from the result, 
to determine the best situation for superconductivity in the Hubbard model 
we have to consider various factors such as the form of the 
pairing interaction and the 
energy/momentum-dependence of Green's functions (lifetime of 
quasi-particles, etc), so that the 
way in which the pairing instability is correlated with 
the lattice structure is a highly nontrivial problem 
which is by no means predictable from the outset.

Here we shall show that 

\noindent (i) $d$-wave instability mediated by 
antiferromagnetic spin fluctuations is stronger than $p$-wave instability
mediated by ferromagnetic spin fluctuations both in 2D and 3D, and

\noindent (ii) pairing instability in 2D is much stronger than that in 3D.

\noindent Thus the `best' situation for the 
spin fluctuation mediated pairing is suggested to be 
the 2D case with dominant antiferromagnetic fluctuations 
as far as the single-band Hubbard model on ordinary lattices are concerned.

\section{Formulation}
\subsection{Model Hamiltonian}
We consider the single-band Hubbard model, 
\begin{equation}
{\cal H}=\sum_{\langle i,j \rangle \sigma} 
t_{ij}c^{\dagger}_{i\sigma}c_{j\sigma}
+U\sum_i n_{i \uparrow}n_{i \downarrow},
\end{equation}
where $c^{\dagger}_{i\sigma}$ creates an electron at the $i$-th site
with spin $\sigma$, $n_{i\sigma}\equiv c^{\dagger}_{i\sigma}c_{i\sigma}$ 
is the number operator.  
We consider the transfer between second-nearest neighbors, $t_{ij}=t'$, 
along with $t (=1$ hereafter) for nearest neighbors.  

The energy dispersions for square and triangular lattices are,
\begin{eqnarray}
\label{SCDIS}
\varepsilon^{\Box}({\bf k})&=&2t
\sum_{i=1}^2\cos(k_i)+4t'\cos(k_1)\cos(k_2),\\
\label{TRIDIS}
\varepsilon^{\triangle}({\bf k})&=&2t
\sum_{i=1}^2\cos(k_i)+2t\cos(k_1+k_2),
\end{eqnarray}
respectively, 
where we take a square Brillouin zone for the latter case as well 
by inserting diagonal transfers in the square lattice.

The dispersions for SC, BCC and FCC lattices are given as
\begin{eqnarray}
\varepsilon^{\rm SC}({\bf k})&=&2t\sum_{i=1}^3 \cos k_i +
4t'\sum_{i<j}\cos k_i \cos k_j, \\
\varepsilon^{\rm BCC}({\bf k})&=&8t
\cos k_1 \cos k_2 \cos k_3
+2t'\sum_{i=1}^3 \cos(2k_i),\\
\label{BCCDIS}
\varepsilon^{\rm FCC}({\bf k})&=&4t\sum_{i<j}
\cos k_i \cos k_j+2t'\sum_{i=1}^{3}\cos(2k_i),
\label{FCCDIS}
\end{eqnarray}
respectively, where $(k_1,k_2,k_3) \equiv (k_x,k_y,k_z)$.  
To facilitate the calculation here we take a cubic Brillouin zone 
($-\pi <k_i\leq \pi$) by considering a four(two) equivalent, 
interpenetrating BCC(FCC) lattices.  

The density of states for the non-interacting case is displayed in 
Fig. \ref{DOS} for the square lattice with $t'=0.0, 0.5$, 
triangular lattice, SC lattice with $t'=0.0, 0.1, 0.3$, 
FCC lattice with $t'=0.0, 0.5$ and BCC lattice with $t'=0.0, 0.1$.

\begin{figure}
\epsfxsize=6cm
\caption{
The density of non-interacting states for
(a) square lattice with $t'=0.0$ (solid line), $t'=0.5$ (dashed line),
(b) triangular lattice,
(c) SC lattice $t'=0.0$ (solid line), $t'=0.1$ (dashed line),
$t'=0.3$ (long dashed line),
(d) BCC lattice with $t'=0.0$ (solid line), $t'=0.1$ (dashed line), and
(e) FCC lattice with $t'=0.0$ (solid line), $t'=0.5$ (dashed line).
}
\label{DOS}
\end{figure}

\subsection{Method}
The FLEX, introduced by Bickers {\it et al.}\cite{FLEX1,FLEX2}, 
starts from a set of skeleton diagrams 
for the thermodynamic potential, $\Phi$, introduced by 
Luttinger and Ward.  $\Phi$ is a functional of Green's 
function, $G$, and a ($k$-dependent) self energy can be computed 
by a functional derivative of $\Phi$ with $G$.  
For the calculation of $\Phi$ 
the idea of Baym and Kadanoff\cite{Baym1,Baym2} of 
taking an important series of diagrams is employed. 
Hence the FLEX approximation is a self-consistent perturbation
approximation with respect to on-site interaction $U$.
The FLEX approximation is suitable for the analysis of Fermi liquid
with strong spin fluctuations.

The self energy is obtained as 
\begin{eqnarray}
\label{Chi}
\Sigma(k)=\frac{1}{N}\sum_{q} G(k-q)V^{(1)}(q),
\end{eqnarray}
in which RPA-type bubble and ladder diagrams are collected for the 
interaction,
\begin{eqnarray}
V^{(1)}(q)&=&\frac{1}{2}U^2\chi_{\rm irr}(q) 
\left[ \frac{1}{1+U\chi_{\rm irr}(q)} \right] \nonumber\\ 
&&+\frac{3}{2}U^2\chi_{\rm irr}(q) 
\left[ \frac{1}{1-U\chi_{\rm irr}(q)} \right] 
-U^2\chi_{\rm irr}(q) \nonumber
\end{eqnarray}
with
\[
\chi_{\rm irr}(q)=-\frac{1}{N}\sum_k G(k+q)G(k).
\]
Here we have denoted $q\equiv ({\bf q},i\epsilon_\nu)$ and 
$k\equiv ({\bf k},i\omega_n)$, 
$\epsilon_\nu=2\pi \nu T$ is the Matsubara frequency for bosons 
while $\omega_n=(2n-1)\pi T$ for fermions, and
$N$ is the total number of sites.
For simplicity, we neglect the diagrams in the particle-particle channel.

The Dyson equation is written as
\begin{equation}
{G({\bf k},\omega_n)}^{-1} = {G^0({\bf k},\omega_n)}^{-1}
-\Sigma({\bf k},\omega_n),
\label{Dyson}
\end{equation}
where $G^0$ is the bare Green's function,
\begin{equation}
{G^0({\bf k},\omega_n)}
 = \frac{1}{{\rm i}\omega_n+\mu-\varepsilon_{\rm k}^0},
\end{equation}
with $\varepsilon_{\rm k}^0$ being the energy of a free electron.
We have solved the equations (\ref{Chi}) $\sim$ (\ref{Dyson}) by 
setting the chemical potential $\mu$ so as to fix the density of electrons.

To obtain $T_{\rm c}$, we solve, with the power method, the eigenvalue 
({\'E}liashberg) equation, 
\begin{eqnarray}
\lambda\Sigma^{(2)}(k)&=&\frac{T}{N}
\sum_{k'}
\Sigma^{(2)}(k')|G(k')|^2 V^{(2)}(k,k'),
\label{eliash}
\end{eqnarray}
where $G(k)$ the dressed Green's function, and $\Sigma^{(2)}(k)$ 
the anomalous self energy, and 
$T=T_{\rm c}$ corresponds to the point at which 
the maximum eigenvalue $\lambda_{\rm Max}$ reaches unity.

The interaction $V^{(2)}$ originates from the transverse spin fluctuations, 
longitudinal spin fluctuations and charge fluctuations, namely,
\begin{eqnarray*}
V^{(2)}(k,k')&=&-U^2\left[\frac{1}{2}\chi_{\rm ch}(k-k') \right.\\
&&\left. -\frac{1}{2}\chi^{zz}(k-k')+\chi^{\pm}(k+k')\right]\\
&=&
- \left[ \frac{U^3\chi_{\rm irr}^2(k-k')}{1-U^2\chi_{\rm irr}^2(k-k')} \right]
- \left[ \frac{U^2\chi_{\rm irr}(k+k')}{1-U\chi_{\rm irr}(k+k')} \right] 
\end{eqnarray*}
where $\chi_{\rm ch}$ is the charge susceptibility, 
while $\chi^{zz} (\chi^{\pm})$ is 
the longitudinal (transverse) spin susceptibility in the 
RPA form where the dressed Green's function is used.

Since we have $\Sigma^{(2)}(k)=\Sigma^{(2)}(-k)$ for the spin-singlet pairing, 
whereas $\Sigma^{(2)}(k)=-\Sigma^{(2)}(-k)$ for the spin-triplet pairing,
$V^{(2)}(k,k')$ becomes a function of $k-k'=q$ with
\begin{eqnarray}
V^{(2)}(q)=
-\frac{3}{2} \left[ \frac{U^2\chi_{\rm irr}(q)}{1-U\chi_{\rm irr}(q)} \right]
+\frac{1}{2} \left[ \frac{U^2\chi_{\rm irr}(q)}{1+U\chi_{\rm irr}(q)} \right]
\label{pair1}
\end{eqnarray}
for the singlet pairing and
\begin{eqnarray}
V^{(2)}(q)=\frac{1}{2} \left[ \frac{U^2\chi_{\rm irr}(q)}{1-U\chi_{\rm irr}(q)} \right]
+\frac{1}{2} \left[ \frac{U^2\chi_{\rm irr}(q)}{1+U\chi_{\rm irr}(q)} \right] 
\label{pair2}
\end{eqnarray}
for the triplet pairing. 
We take $N=64^2$ sites with $n_c=2048$ Matsubara frequencies for 2D
square lattice and with $n_c=1024$ Matsubara frequencies for 2D
triangular lattice, and $N=32^3$ with $n_c=1024$ for 3D.

\section{Results}
\subsection{Square lattice with antiferromagnetic spin fluctuations}
Let us start with the case of nearly half-filled square lattice, which 
has strong antiferromagnetic fluctuations.
In this case, $T_{\rm c} \sim 0.02$ for 
$d$-wave pairing should be obtained within the FLEX approximation
as mentioned in the Introduction.  
We first present the result for 
$t'=0.0$, $n=0.85$ (0.15 holes doped) and $U=4$, which will 
serve as a reference for other lattices.

In Fig. \ref{sl-gr}, we show 
$|G({\bf k}, {\rm i}\pi k_{\rm B} T)|^2$, a quantity which appears in 
the right-hand side of the {\'E}liashberg equation (\ref{eliash}).  
We can see that $|G|^2$ takes large values 
($\sim 8.0$) around the Fermi surface.  
Note that the smaller the self energy correction,
the peak of $|G({\bf k}, {\rm i}\pi k_{\rm B} T)|^2$ becomes larger 
and the Fermi-liquid picture becomes more valid.  
Also, a large $|G({\bf k}, {\rm i}\pi k_{\rm B} T)|^2$ favors 
superconductivity as we can see from 
the {\'E}liashberg equation (\ref{eliash}).

In Fig. \ref{sl-chi}, we plot the susceptibility in the RPA form, 
$\chi({\bf k},0)=\chi_{\rm irr}/(1-U\chi_{\rm irr})$, 
as a function of the wave number 
for $T=0.03$. Dominant antiferromagnetic spin fluctuations are seen 
as $\chi$ peaked around ${\bf k}=(\pi,\pi)$.  
A large peak in $\chi({\bf k},0)$ should imply 
a large pairing interaction as
seen in the {\'E}liashberg equation (\ref{eliash}).  
The spread of $\chi({\bf k},0)$ around the peak 
in the momentum space is also important as we shall come back later, 
because it
measures the fraction of effective channels that contribute 
to $V^{(2)}$ in the {\'E}liashberg equation (\ref{eliash}). 

We also plot in Fig. \ref{sl-chi} Im$\chi({\bf k}_{\rm Max},\omega)$ 
against $\omega$, 
where ${\bf k}_{\rm Max}$ is the momentum for which $\chi({\bf k},0)$ 
becomes maximum and we have normalized 
Im$\chi({\bf k}_{\rm Max},\omega)$ with the maximum value.  
We have obtained the dependence on real $\omega$ with an 
analytic continuation in the Pad{\'e} approximation\cite{Pade}.
The spread of Im$\chi({\bf k}_{\rm Max},\omega)$ around the peak in
the $\omega$ sector is an electronic couterpart to 
the Debye frequency $\omega_D$
in the BCS theory for the electron-phonon system, 
so this quantity is another important factor in the pairing instability.

In Fig. \ref{sl-super}, we plot $\lambda_{\rm Max}$ as a function of 
temperature $T$ (normalized by $t$) 
along with the reciprocal of the peak value of $\chi({\bf k},0)$. 
The pairing instability is measured by how 
$\lambda_{\rm Max}$ is close to unity, while 
$1/\chi \rightarrow 0$ signifies the magnetic ordering. 
$\lambda_{\rm Max}$ is seen to be extrapolated to 
unity at $T\sim 0.02$, 
in accord with previous results\cite{Dahm}.  

Here let us comment on a finite $T_{\rm c}$ in 2D systems. 
As usually done, it is taken to be a measure of $T_{\rm c}$ 
when the layers are stacked with the Josephson coupling. 
However, if we take account of the superconducting fluctuations rigorously,
$T_{\rm c}$ for a purely 2D system must be zero 
according to Mermin's theorem\cite{Mermin,Hohenberg,2Dsupercomm}.
To judge whether a mean field treatment of pairing instabilities 
is adequate will require an evaluation of 
the coherence length of the pairing.  
Although this has been done for the conventional 
phonon mechanism of superconductivity with Gor'kov's argument\cite{gorkov}, 
such an evaluation is rather difficult in the present case of an electron 
mechanism, since the pairing potential strongly depends on 
frequency and wave-number.

We can on the other hand study the effect
of a weak three dimensionality on the $\lambda_{\rm Max}$ 
within the present formalism.  
Specifically, we introduce an inter-layer hopping, $t_z$.  
Figure \ref{2d3d} shows the results for $\lambda_{\rm Max}$ 
for a 3D anisotropic cubic lattice with the inter-layer hopping 
varied over $0.0<t_z<1.0$ on top of $t_x=t_y=1.0$. 
We can see that weak but finite $t_z<0.3$ does not appreciably 
change the behavior of $\lambda_{\rm Max}$.  
Thus, provided the coherence length in $z$-direction is 
large enough for $t_z \simeq 0.3$ to validate the mean-field treatment, 
$T_{\rm c}$ obtained here for 2D can indeed be used as a measure 
of $T_{\rm c}$ when a weak three dimensionality is present.

\begin{figure}
\epsfxsize=6cm
\caption{
The squared absolute value of the Green 
function for the smallest Matsubara frequency, 
${\rm i}\omega_n={\rm i}\pi k_{\rm B} T$, against 
wave number for the Hubbard model on a square lattice
with $t'=0$, $n=0.85$ and $U=4$.
}
\label{sl-gr}
\end{figure}

\begin{figure}
\epsfxsize=6cm

\epsfxsize=6cm
\caption{
(a)${\rm Im}\chi({\bf k}_{\rm Max},\omega)$ 
(normalized by its maximum value)
as a function of $\omega/t$ 
for the Hubbard model on a square lattice
with $t'=0.0$, $n=0.85$, $U=4$ and $T=0.03$.
(b)$\chi({\bf k},0)$ 
as a function of wave number for the Hubbard model 
on a square lattice with $t'=0.0$, $n=0.85$, $T=0.03$ and $U=4$.
}
\label{sl-chi}
\end{figure}

\begin{figure}
\epsfxsize=6cm
\caption{
The maximum eigenvalue of the {\'E}liashberg equation 
(solid lines) 
and the reciprocal of the value of $\chi$ (dashed lines) at the 
antiferromagnetic peak 
against temperature for 
the Hubbard model on a square lattice with $t'=0$, $n=0.85$ and $U=4$.
}
\label{sl-super}
\end{figure}

\begin{figure}
\epsfxsize=6cm
\caption{
The maximum eigenvalue of the {\'E}liashberg equation 
against temperature for 
the Hubbard model on an anisotropic 
cubic lattice with $t_x=t_y=1$ and $t_z=0.0\sim1.0$,
$n=0.85$ and $U=4$.
}
\label{2d3d}
\end{figure}

\subsection{$t$-$t'$ square lattice with ferromagnetic spin fluctuations}
We move on to
a 2D case with dominant ferromagnetic spin fluctuations, where 
triplet pairing is expected. 
The situation for which the ferromagnetic
fluctuations become dominant
has extensively been investigated for the Hubbard model
with various approaches, and one guiding principle is that 
a large density of states at the Fermi level
located near the bottom of the band should 
favor ferromagnetism for sufficiently strong electron-electron repulsion.

For the square lattice, such a situation may be realized 
for relatively large $t'(\simeq 0.5)$ 
for dilute electron densities.  
In this case, the divergence (van Hove singularity) in the density of 
states, which is at the center and has a functional form $|{\rm ln} 
E|$ for $t'=0$, shifts toward the band bottom 
with $D(E) \sim 1/\sqrt{(2+E)}|{\rm ln}(2+E)|$ 
for $t'=0.5$, 
so that the density of states at the Fermi level 
becomes large for the dilute case, favoring ferromagnetism\cite{Namaizawa}.  
It has in fact been shown from quantum Monte Carlo studies 
that the ground state is fully spin-polarized for 
$t'=0.47$, $n\sim 0.4$.\cite{Hlubina,Hlubina99}
So we explore the possibility of $p$-wave
instability associated with this ferromagnetism, 
and compare the result with that of $d$-wave instability
associated with the antiferromagnetism discussed in the previous subsection.

First, in Fig. \ref{ttlambda}, we show $\lambda_{\rm Max}$ 
for the density varied over $0.3\leq n \leq 0.6$ and 
$t'$ varied over $0.3 \leq t' \leq 0.6$ for $U=4, 6$ with $T=0.03$.
We can see that $\lambda_{\rm Max}$ becomes largest for $n=0.3$ and 
$t'=0.5$ for both $U=4$ and 6, so we take these parameter sets.

In Fig. \ref{tt-chi}, we plot
$\chi({\bf k},0)$ for $U=4$ as a function of momentum
along with Im$\chi({\bf k}_{\rm Max}, \omega)$
as a function of $\omega$.
The peak of $\chi({\bf k},0)$ is indeed located at 
the $\Gamma$ point $({\bf k}=(0,0))$ indicating ferromagnetic fluctuations.  
The frequency spread is similar to the case of the nearly 
half-filled $t'=0$ square lattice.  
In Fig.\ref{tt-super}, we plot 
$\lambda_{\rm Max}$ as a function of $T$.
We can see that $\lambda_{\rm Max}$ is much smaller than 
that in the antiferromagnetic case, Fig.\ref{sl-super}. 

\begin{figure}
\epsfxsize=6cm
\caption{
The maximum eigenvalue of the
{\'E}liashberg equation, 
$\lambda_{\rm Max}$, for a square lattice with the density 
varied over $0.3 \leq n \leq 0.6$ and 
$t'$ varied over $0.3 \leq t' \leq 0.6$ for $U=4,6$ 
with $T=0.03$.
}
\label{ttlambda}
\end{figure}

\begin{figure}
\epsfxsize=6cm

\epsfxsize=6cm
\caption{
(a)${\rm Im}\chi({\bf k}_{\rm Max},\omega)$ 
(normalized by its maximum value)
as a function of $\omega/t$
for the Hubbard model on a square lattice
with $t'=0.5$, $n=0.3$, $U=4$ and $T=0.03$.
(b)$\chi({\bf k},0)$ 
as a function of wave number for the Hubbard model 
on a square lattice with $t'=0.5$, $n=0.3$, $T=0.03$ and $U=4$.
}
\label{tt-chi}
\end{figure}

\begin{figure}
\epsfxsize=6cm
\caption{
A plot similar to Fig. \protect\ref{sl-super} 
for the Hubbard model on a square lattice
with $t'=0.5$, $n=0.3$ and $U=4$.
The dashed line denotes the reciprocal of 
the ferromagnetic peak in $\chi$.
}
\label{tt-super}
\end{figure}

\begin{figure}
\epsfxsize=6cm
\caption{
A plot similar to Fig. \protect\ref{sl-gr} 
for the Hubbard model on a square lattice
with $t'=0.5$, $n=0.3$ and $U=4$.
}
\label{tt-gr}
\end{figure}

\subsection{Why is p-pairing weaker than d-pairing?}
The present result that the p-pairing has a lower $T_{\rm c}$ 
contrasts with a naive expectation from the BCS picture, 
in which the $T_{\rm c}$ should be high for a large density of states 
at the Fermi level. We may trace back the reason 
why this does not apply as follows.

First, if we look at the dominant 
($\propto 1/[1-U\chi_{\rm irr}(q)]$) term in 
the pairing potential $V^{(2)}$ itself 
in eqs. (\ref{pair1}) and (\ref{pair2}), 
the triplet pairing interaction is only one-third of 
the singlet pairing interaction.  So this should be one reason.  

On the other hand, the large density of states, namely the 
flatness of the band around the Fermi level, 
is reflected to the fact that 
$|G|^2$ (Fig.\ref{tt-gr}) forms an almost flat plateau 
in a large portion of the Brillouin zone.
If the maximum value of $|G|^2$ is large enough to compensate 
the disadvantage of the one-third $V^{(2)}$, a large $\lambda$ may emerge.
However, we can see that $|G|^2$ 
is much smaller than that in the antiferromagnetic case (Fig.\ref{sl-gr}),
which implies that the self energy correction is large.  
Even when we take a larger repulsion $U$ to increase the 
triplet pairing attraction (i.e., to increase the susceptibility), 
this makes the self-energy correction 
even greater, resulting in only a slight change in $\lambda$.

\subsection{2D triangular lattice}
Next, we discuss the case of 2D triangular lattice.
As mentioned in the Introduction, the half-filled case with 
dominant antiferromagnetic fluctuations 
has already been discussed by a number of authors.
\cite{KinoKontani,KonMori,Schmalian,VojtaDaggoto,Jujo1,Jujo2}

Here, we focus on the quarter-filled isotropic triangular lattice, 
where we can expect ferromagnetic fluctuations, and hence 
$p$-wave pairing.  As we can see in Fig. \ref{DOS},
the density of states for the triangular lattice has a sharp peak 
(with $D(E) \sim |{\rm ln}(E+2)|$) 
near the bottom of the band, so
a dominant ferromagnetic fluctuation is expected
if the Fermi level is located at the peak.
Hanish {\it et al.}\cite{Hanisch} 
studied the instability of the fully-ferromagnetic state 
of Hubbard model on the triangular lattice for large $U$ and
concluded that
the ferromagnetic state is stable for $n\simeq 0.5$ (quarter-filled).

In this situation, we have calculated $\lambda_{\rm Max}$
for $T=0.03$, $U=4,8,12$.
Since $\lambda_{\rm Max}$ takes
similar values for $U=4, 8, 12$, we take $U=8$.

In Fig. \ref{tri-chi}, we plot 
the wave-number dependence of $\chi({\bf k})$ 
and frequency dependence of Im$\chi({\bf k}_{\rm Max},\omega)$.
We can see that the peak of $\chi({\bf k},0)$ is located around
the $\Gamma$ point.
In Fig. \ref{tri-super}, we plot $\lambda_{\rm Max}$ along with 
the reciprocal of the peak value of $\chi({\bf k},0)$
as a function of $T$.  We can see that 
the $p$-wave instability here is again much weaker than
the $d$-wave instability for the nearly half-filled square lattice 
(Fig. \ref{sl-super}).  
The one-third factor discussed in the previous subsection 
should again be one factor. 
The second factor is also involved, namely, the maximum value of 
$U^2 |G({\bf k},i k_{\rm B} T)|^2$ is small ($\sim 1.0\times 8^2$, see 
Fig. \ref{tri-gr}) compared to that in Fig. \ref{sl-gr} 
($\sim 8.0 \times 4^2$), so the quasi-particles are short-lived.

We note that the ferromagnetic fluctuation in the triangular case is much 
weaker than in the square lattice with large $t'$ 
(compare Figs. \ref{tri-chi}(b) and \ref{tt-chi}(b)),
although we have taken a larger $U$ here. 
This may be because the peak in the density of states 
in the triangular lattice 
is situated slightly above the bottom
(compare Figs.\ref{DOS} (a) and (b)). 
Onoda and present authors have suggested in 
ref.\cite{AriOnoKuAo} that closer the peak of the density of 
states is to the band bottom,
the stronger the ferromagnetic fluctuations tend to be.

Despite this difference in ferromagnetic fluctuation, the 
$\lambda_{\rm Max}$ in the present case 
is comparable to that in the square lattice with $t'$.  
This may be because the frequency spread in the present case 
is quite large as seen in Fig.\ref{tri-chi}(a).

\begin{figure}
\epsfxsize=6cm

\epsfxsize=6cm
\caption{
(a)${\rm Im}\chi({\bf k}_{\rm Max},\omega)$ 
(normalized by its maximum value)
as a function of $\omega/t$
for the Hubbard model on a triangular lattice
with $n=0.5$, $U=8$ and $T=0.03$.
(b)$\chi({\bf k},0)$ 
as a function of wave number for the Hubbard model 
on a triangular lattice with $n=0.5$, $T=0.03$ and $U=8$.
}
\label{tri-chi}
\end{figure}

\begin{figure}
\epsfxsize=6cm
\caption{
A plot similar to Fig. \protect\ref{sl-super} 
for the Hubbard model on a triangular lattice, with $n=0.5$ and $U=8$.
The dashed line denotes the reciprocal of 
the ferromagnetic peak of $\chi$.
}
\label{tri-super}
\end{figure}

\begin{figure}
\epsfxsize=6cm
\caption{
A plot similar to Fig. \protect\ref{sl-gr} 
for the 2D Hubbard model on triangular lattice
with $n=0.5$, and $U=8$.
}
\label{tri-gr}
\end{figure}

\subsection{3D simple cubic lattice}
Let us now move on to the case of 3D systems.
we first discuss the possibility of $d$-wave pairing
in the nearly half-filled Hubbard model on the 3D SC lattice.
In this case, we find that the $\Gamma_{3}^{+}$ representation 
($\sim x^2-y^2$ etc,
with the gap function $\Delta({\bf k}) 
\sim {\rm cos}k_x - {\rm cos}k_y$ etc) 
of O$_{h}$ group\cite{SigristUeda} has the largest $\lambda_{\rm Max}$, 
so we concentrate on this pairing symmetry.

In Fig. \ref{3dsclambda}, we show $\lambda_{\rm Max}$ for the density 
varied over $0.75 \leq n \leq 0.9$ and 
$t'$ varied over $-0.5 \leq t' \leq +0.4$ for $U=4,6,8,10,12$ 
with $T=0.03$.
We can see that among these parameter sets, $\lambda_{\rm Max}$ 
becomes largest for $n=0.8$, $t'=-0.2\sim -0.3$ and $U=8\sim 10$,
so we take these parameter sets.

In Fig. \ref{sc2}(a), we show the momentum dependence of
$\chi({\bf k},0)$ for $U=8$, $n=0.8$, $t'=-0.3$, $T=0.03$.
We can see that the peak of $\chi({\bf k},0)$
is located around the K point $(\pi,\pi,\pi)$ as expected 
for the antiferromagnetism.

In Fig. \ref{3dsc-super}, we plot $\lambda_{\rm Max}$ along with 
the reciprocal of the peak value of $\chi({\bf k},0)$ as a 
function of $T$ for $t'=-0.2,-0.3$ ,$U=8$ and $n=0.8$.
We see that $\lambda_{\rm Max}$ does not become very close to unity 
in the range calculated here.  
For $T<0.02$, the result obtained for $N=32^3$ and $n_c=1024$ 
is not convergent enough with respect to 
the system size and the number of Matsubara frequencies, 
so we show the result for a larger $n_c=2048$ in the inset, 
which suggests that
$\lambda_{\rm Max}$ tends to increase with $n_c$.
We have also performed a calculation for $N=16^3$, and 
found that $\lambda_{\rm Max}$ also increases with $N$,
so a finite $T_{\rm c}$ ($<0.01$) may be obtained at least
for $t'=-0.3$, $U=8$, $n=0.8$ in the limit of large $N$ and $n_c$.
At any event, $T_{\rm c}$ is significantly smaller than in the square lattice
with strong antiferromagnetic fluctuations (Fig. \ref{sl-super}).  

\begin{figure}
\epsfxsize=6cm
\caption{
A plot similar to Fig. \protect\ref{ttlambda} 
for an SC lattice, with the density 
varied over $0.75 \leq n \leq 0.9$ and 
$t'$ varied over $-0.5 \leq t' \leq 0.4$ for $U=4,6,8,10,12$ 
with $T=0.03$.
Where the curves are truncated the system becomes antiferromagnetic 
(i.e., $U\chi_{\rm irr}$ becomes unity).  
}
\label{3dsclambda}
\end{figure}

\begin{figure}
\epsfxsize=8cm

\epsfxsize=6cm
\caption{
(a) $\chi({\bf k},0)$ 
as a function of wave number for the Hubbard model 
on an SC lattice with $t'=-0.2$, $n=0.8$, $T=0.03$ and $U=8$.
(b)${\rm Im}\chi({\bf k}_{\rm Max},\omega)$ 
(normalized by its maximum value)
as a function of $\omega/t$
for the Hubbard model on an SC lattice(dashed line)
with $t'=-0.2$, $n=0.8$, $U=8$, as compared with 
the Hubbard model on a square lattice(solid line)
with $t'=0$, $n=0.85$, $U=4$ for $T=0.03$ 
(left frame, where the 2D and 3D results almost overlap) 
or $T=0.04$ (right frame).  
}
\label{sc2}
\end{figure}

\begin{figure}
\epsfxsize=6cm
\caption{
A plot similar to Fig. \protect\ref{sl-super} 
for the Hubbard model on an SC lattice with $t'=-0.2,-0.3$, $n=0.8$ and $U=8$.
The inset shows the results for a larger 
number (=2048) of Matsubara frequencies for $t'=-0.3$.
}
\label{3dsc-super}
\end{figure}

\begin{figure}
\epsfxsize=5cm
\caption{
A plot for the squared Green function against $k_x$ and $k_y$ 
with $k_z=0, \pi/2, \pi$ for the Hubbard model on an SC lattice
with $t'=-0.2$, $n=0.8$ $U=8$, $T=0.03$. 
}
\label{3dscgr}
\end{figure}

\subsection{Why is $d$-pairing stronger in 2D than in 3D?}
So the $d$-wave instability is 
decidedly stronger in 2D than in 3D as far as the square 
and simple cubic lattices are concernded, and the 
question is: what are physical reasons for that.  
We can pinpoint the origin by looking at 
the factors involved in the {\'E}liashberg equation (\ref{eliash}), 
i.e., (a) the factor $U^2 |G|^2$, (b) the summation over the 
frequency, and (c) the summation over the momentum. 
In addition, the factor $V^{(2)}$ is of course important in the equation, 
but in the following we compare 2D and 3D in the situation where 
the maximum value of $\chi({\bf k},0)$ (that determines $V^{(2)}$) 
is similar between the two cases to concentrate on the factors (a)(b)(c).  

In Fig. \ref{3dscgr} we plot $|G|^2$ in 3D for $k_z=0,\pi/2,\pi$ 
as a function of $k_x$ and $k_y$ 
for $U=8$, $n=0.8$, $t'=-0.3$, $T=0.03$.
We can see that the maximum value of $U^2 |G|^2$ 
is greater in 3D than in 2D. 

Fig. \ref{sc2}(b) 
displays ${\rm Im}\chi({\bf k}_{\rm Max},\omega)$ 
as a function of $\omega$.
The figure compares the SC lattice ($t'=-0.2, n=0.8, U=8$) 
with a typical square lattice with $t'=0$, 
$n=0.85$ and $U=4$ having a similar magnitude of $\chi$ at $T=0.03$.  
We can see that the frequency spread of 
${\rm Im}\chi({\bf k}_{\rm Max},\omega)$
is similar between 3D and 2D, so the factors (a)(b) 
can be excluded from the reason for the 2D-3D difference.
Note that if the frequency spread of the susceptibility is
scaled not by $t$ but 
by the {\it band width}, as Nakamura {\it et al}\cite{Nakamura}
have assumed, $\lambda_{\rm Max}$ would have become larger. 

If we turn to the momentum sector in the susceptibility, 
$\chi({\bf k},0)$, Fig. \ref{sc2}(a) shows that 
the width, $a$, of the $\chi({\bf k},0)$ peak in each 
momentum component 
is similar to those in 2D displayed in Fig.\ref{sl-chi}, 
where the main contribution of $V^{(2)}$ to $\lambda$ is confined 
around $(\pi,\pi)$ in 2D or $(\pi,\pi,\pi)$ in 3D.
Since the right-hand side of the {\'E}liashberg equation (\ref{eliash}) 
is normalized by $N \propto L^{D}$ 
with $L$ being the linear dimension of the system, 
$\lambda$ is proportional to $(a/L)^D$, 
so that if $a$ has similar values between 2D and 3D 
we end up with a smaller $\lambda$ in 3D than that in 2D.  
So we can conclude that this is the main reason why the 2D case
is more favorable than the 3D case.  

\subsection{BCC lattice}
Let us turn to the BCC lattice.
For BCC lattice, the density of states diverges around the 
center with $D(E) \sim [{\rm ln}(E)]^2$, and 
the antiferromagnetic fluctuation is dominant
near half-filling\cite{AriOnoKuAo}, so we focus on the possibility
of $d$-wave superconductivity.
For the d-wave, we found that $\Gamma_5^+$ representation 
($\sim xy$ etc, 
with the gap function $\Delta({\bf k}) 
\sim {\rm cos}(k_x+k_y+k_z)-{\rm cos}(k_x+k_y-k_z)$, etc) 
of $O_h$ group has the largest $\lambda_{\rm Max}$.

In Fig. \ref{bcclambda}, we show $\lambda_{\rm Max}$
for the density varied over $0.75 \leq n \leq 0.9$ and 
$t'$ varied over $-0.4 \leq t' \leq 0.4$ for $U=4,6,8$ 
with $T=0.03$.
Antiferromagnetic spin fluctuations are much stronger in BCC lattice 
than in the SC lattice as pointed out in ref. \cite{AriOnoKuAo}.  
In fact, in this figure, the truncated curves for $\lambda_{\rm Max}$ 
means that the system becomes antiferromagnetic 
(i.e., $U\chi_{\rm 0}$ becomes unity) there.  
Hereafter, we focus on the case of $U=6$, $t'=0.1$, $n=0.75$.

In Fig. \ref{bcc-chi}(a), we plot 
the momentum dependence of $\chi({\bf k},0)$. 
We can see that the peak of $\chi({\bf k},0)$ is 
located around $K$-point (${\bf k} = 
(\pi,\pi,\pi))$. 
In Fig. \ref{bcc-chi}(b), we plot 
the frequency dependence of Im$\chi({\bf k}_{\rm Max},\omega)$
which shows that the frequency spread for the BCC lattice is 
similar to the case of the square lattice or the SC lattice.
In Fig. \ref{bcc-super}, we plot $\lambda_{\rm Max}$ for $d$-wave
pairing as a function of $T$. 
$\lambda_{\rm Max}$ is again much smaller than that for 
the nearly half-filled square lattice, and is even smaller than 
that for the SC lattice.
The reason for the former is mainly due to the $(a/L)^D$ factor 
discussed above, 
while the reason for the latter is because 
the maximum in $U^2 |G|^2$ in BCC lattice
is smaller (see Fig. \ref{bcc-gr}) 
than that in SC lattice (Fig. \ref{3dscgr}).

\begin{figure}
\epsfxsize=6cm
\caption{
A plot similar to Fig. \protect\ref{ttlambda} 
for a BCC lattice, with the density 
varied over $0.75 \leq n \leq 0.85$ and 
$t'$ varied over $-0.4 \leq t' \leq 0.4$ for $U=4,6,8$ 
with $T=0.03$.
Where the curves are truncated the system becomes antiferromagnetic 
(i.e., $U\chi_{\rm irr}$ becomes unity).  
}
\label{bcclambda}
\end{figure}

\begin{figure}
\epsfxsize=6cm

\epsfxsize=6cm
\caption{
(a) $\chi({\bf k},0)$ 
as a function of wave number for the Hubbard model 
on a BCC lattice with $t'=0.1$, $n=0.75$, $T=0.03$ and $U=6$.
Note that if we take the energy dispersion (\protect\ref{BCCDIS}), 
$(\pi,\pi,\pi)$ and $(\pi,0,0)$ are equivalent. 
(b)${\rm Im}\chi({\bf k}_{\rm Max},\omega)$ (normalized by its maximum value)
as a function of $\omega/t$ for the Hubbard model on a BCC lattice
for $t'=0.1$, $n=0.75$, $U=6$ with $T=0.03$ (left frame) or 
$T=0.04$ (right).  
}
\label{bcc-chi}
\end{figure}

\begin{figure}
\epsfxsize=6cm
\caption{
The maximum eigenvalue of the {\'E}liashberg equation 
(solid lines) 
and the reciprocal of the antiferromagnetic peak in $\chi$ 
(dashed lines) 
against temperature for 
the Hubbard model on a BCC lattice with $t'=-0.1$, $n=0.75$ and $U=6$.
}
\label{bcc-super}
\end{figure}

\begin{figure}
\epsfxsize=5cm
\caption{
A plot for the squared Green function 
against $k_x$ and $k_y$ with $k_z=0, \pi/2, \pi$ 
for the Hubbard model on a BCC lattice
with $t'=0.1$, $n=0.75$, $U=6$ and $T=0.03$. 
}
\label{bcc-gr}
\end{figure}

\subsection{FCC lattice}
We finally come to the FCC lattice.
The density of states of FCC lattice diverges at the bottom
of the band with $D(E) \sim |{\rm ln} (E+4)|$ 
for $t'=0$ and $\sim 1/\sqrt{E+3}$ for $t'=0.5 0$, 
so we can expect large 
ferromagnetic spin fluctuations for low densities of electrons.
In fact, the possibility of ferromagnetic ground
state has been discussed using
various approaches\cite{Hanisch,Ulmke,AriOnoKuAo}.
According to our previous study\cite{AriOnoKuAo},
the ferromagnetic spin fluctuation is most dominant
for $n\sim 0.2$ and $t'\simeq 0.5$ in the weak coupling regime.
Here, we focus on the possibility of $p$-wave pairing 
at low densities.
In this case, we found that the case of 
the gap function $\Delta({\bf k})\sim \sin(k_x+k_y)$ 
has the largest $\lambda_{\rm Max}$.
(Since the Hubbard model has SU(2) symmetry, 
the $T_{\rm c}$ does not depend on the direction of the d-vector,
so that we may concentrate on the wave-number dependence.)


In Fig. \ref{fcclambda}, we show the maximum eigenvalue
of the {\' E}liashberg equation 
for the density varied over $0.2 \leq n \leq 0.4$ and 
$t'$ varied over $0.0 \leq t' \leq 0.6$ for $U=2,4,6,8$ 
with $T=0.03$.  We can see that $\lambda_{\rm Max}$ 
is much smaller than unity even around $t'\simeq 0.5$.

To probe the origin of this behavior, we 
take the case of $U=2$, $t'=0.5$, and $n=0.3$, 
which is also convenient 
becase the maximum value of $\chi({\bf k},0)$ (see below) 
in this case takes a similar value as in the square lattice with $t'=0.5$ 
(Fig.\ref{tt-chi}(b)), which facilitates the comparison between the two cases. 
In Fig. \ref{fcc-chi}(b), we show the momentum dependence of 
$\chi({\bf k},0)$.  
We can see that the peak is indeed located around 
the ferromagnetic point ($\Gamma$).  

In Fig. \ref{fcc-super}, we plot $\lambda_{\rm Max}$ along with 
the reciprocal of the peak value of $\chi({\bf k},0)$ as a 
function of $T$. $\lambda_{\rm Max}$ in this case is much smaller 
than that for the $d$-wave pairing in the SC lattice.  
This is again mainly due to
the one-third $V^{(2)}$. 
$\lambda_{\rm Max}$ is smaller even when compared with that for 
the $p$-wave pairing in the square lattice with $t'=0.5$,
although the maximum value of $\chi({\bf k},0)$, 
the frequency spread of 
${\rm Im}\chi({\bf k}_{\rm Max},\omega)$ (Fig. \ref{fcc-chi}(a)),
and $U^2 |G|^2$ (see Fig. \ref{fcc-gr}) all take similar values 
between the two. In fact, the reason for this 2D-3D 
discrepancy can again be traced back to the $(a/L)^D$ factor
discussed in the previous subsection.

\begin{figure}
\epsfxsize=6cm
\caption{
A plot similar to Fig. \protect\ref{ttlambda} 
for an FCC lattice, with the density 
varied over $0.2 \leq n \leq 0.4$ and 
$t'$ varied over $0.0 \leq t' \leq 0.6$ for $U=2,4,6,8$ 
with $T=0.03$.
Where the curves are truncated the system becomes ferromagnetic 
(i.e., $U\chi_{\rm irr}$ becomes unity).  
}
\label{fcclambda}
\end{figure}

\begin{figure}
\epsfxsize=6cm

\epsfxsize=6cm
\caption{
(a)${\rm Im}\chi({\bf k}_{\rm Max},\omega)$ 
(normalized by its maximum value)
as a function of $\omega/t$
for the Hubbard model on an FCC lattice
with $t'=0.5$, $n=0.3$, $U=2$ and $T=0.03$
(b)$\chi({\bf k},0)$ 
as a function of wave number for the Hubbard model 
on an FCC lattice with $t'=0.5$, $n=0.3$, $T=0.03$ and $U=2$.
Note that the $\Gamma$-point (0,0,0) and the $K$-point $(\pi,\pi,\pi)$
are equivalent if we take the dispersion (\protect\ref{FCCDIS}).
}
\label{fcc-chi}
\end{figure}

\begin{figure}
\epsfxsize=6cm
\caption{
The maximum eigenvalue of the {\'E}liashberg equation 
(solid lines) 
and the reciprocal of the ferromagnetic peak of $\chi$ 
(dashed lines) 
in the Hubbard model on an FCC lattice with $t'=0.5$, $n=0.3$ and $U=2$.
}
\label{fcc-super}
\end{figure}

\begin{figure}
\epsfxsize=5cm
\caption{
The squared Green function against $k_x$ and $k_y$ 
with $k_z=0, \pi/2, \pi$ 
for the Hubbard model on an FCC lattice
with $t'=0.5$, $n=0.3$, $U=2$ and $T=0.03$. 
}
\label{fcc-gr}
\end{figure}

\section{Discussions and Summary}
To summarize, we have studied the possibility of spin-fluctuation mediated 
superconductivity in the single-band, repulsive Hubbard model
for the $d$-wave channel on the (a) square lattice,
(b) SC lattice, (c) BCC lattice,
and for the $p$-wave channel on the 
(d) square lattice with large second-nearest neighbor
hopping, (e) triangular lattice, and (vi) FCC lattice.
We have shown that (i) $d$-wave instability mediated by antiferromagnetic
spin fluctuations is stronger than $p$-wave instability
mediated by ferromagnetic spin fluctuations both in 2D and 3D, and
(ii)$d$-wave instability in 2D is much stronger than that in 3D.

We have given the physical reasons why the 
triplet p-pairing is unfavored as 
(i) the pairing interaction $V^{(2)}$ for triplet pairing
is only $1/3$ of those for the singlet pairing, 
(ii) the self energy correction is large (i.e., quansi-particles 
are short-lived).

We have also traced back physical reasons why 
the superconducting instability in 3D is weaker than in 2D: if
the momentum spread of $\chi$ (that determines 
$V^{(2)}$) in each momentum direction 
take similar values in 2D and 3D, it
makes the higher-dimensional 3D disadvantageous because of a structure of the 
{\' E}liashberg equation. 
Thus, our conclusion is that, 
so far as the single-band Hubbard model on ordinary lattices are concerned, 
$d$-wave pairing in 2D square lattice is the ``best" situation for the 
spin fluctuation mediated superconductivity, 
where $T_{\rm c}$ can reach $O(0.001W)$ 
if we measure $T_{\rm c}$ in units of the band width, $W$.  
In this sense, the layer-type cuprates do seem to hit upon the 
right situation.  

However, our conclusion has been obtained 
for the single-band Hubbard model.  
If we look more extensively at 3D superconductors, 
the heavy fermion system, in which the pairing 
is also thought to be meditated by spin fluctuations,
$T_{\rm c}$ has a similar order of magnitude 
$O(0.01W \sim 0.001W)$, 
i.e., $T_{\rm c} \sim 1$ K with $W=$ a few hundred K\cite{Nakamura}.
Since the present result indicates that $T_{\rm c}$ 
in the 3D Hubbard model should be much smaller than that for the 
2D Hubbard model, we can envisage that the heavy fermion system 
exploits a more favorable situation than in the single-band 
Hubbard model where the frequency and/or momentum spreads in 
$\chi({\bf k},\omega)$ are larger than those in the 3D Hubbard model.  
In fact, the standard models, such as the multiband periodic Anderson model, 
employed to describe the heavy fermion system are 
more complicated than the single-band Hubbard model, and 
the frequency/momentum spreads in $\chi$ 
in such a model should be large enough to explain the superconductivity 
in the heavy fermion systems in the present context.
It is an appealing future problem to explore how this can be so.

\section{Acknowledgment}
We would like to thank Professor K. Ueda and Dr H. Kontani for illuminating 
discussions.
R.A. would like to thank Dr S. Koikegami for discussions on the FLEX.
R.A. is supported by JSPS, while 
K.K. acknowledges a Grant-in-Aid for Scientific
Research from the Ministry of Education of Japan.
Numerical calculations were performed at the Supercomputer Center,
ISSP, University of Tokyo.

\end{document}